\documentclass[preprint]{aastex}

\def\aap{\ {A\&A}\ }

\def\aj{\ {AJ}\ }

\def\apj{\ {ApJ}\ }
\def\apjl{\ {ApJL}\ }
\def\apjs{\ {ApJS}\ }
\def\apss{\ {Ap\&SS}\ }
\def\araa{\ {ARA\&A}\ }

\def\mnras{\ {MNRAS}\ }

\def\nat{\ {Nat}\ }

\def\pasp{\ {PASP}\ }

\newcommand{\MSun}{\mbox{${\rm M}_\odot$}}
\newcommand{\ZSun}{\mbox{$Z_\odot$}}
\newcommand{\LSun}{\mbox{$L_\odot$}}

\def\apgt{\ {\raise-.5ex\hbox{$\buildrel>\over\sim$}}\ }
\def\aplt{\ {\raise-.5ex\hbox{$\buildrel<\over\sim$}}\ }
\def\lteq{\ {\raise-.5ex\hbox{$\buildrel<\over-$}}\ }

\begin{document}

\title{Are Super-Luminous Supernovae and Long GRBs the products of dynamical processes in young dense star clusters?}
\author{E.P.J. van den Heuvel}
\affil{Astronomical Institute ‘Anton Pannekoek’, University of Amsterdam, 
P.O. Box 94249, 1090 GE, Amsterdam, The Netherlands} 
\affil{Institute for Theoretical Physics, UCSB, Santa Barbara, CA 93106-4030}

\and 

\author{S.F. Portegies Zwart}
\affil{Leiden Observatory, Leiden University, PO Box 9513, 2300 RA, Leiden, 
The Netherlands}

\begin{abstract}
Super Luminous supernovae (SLSN) occur almost exclusively in small
galaxies (SMC/LMC-like or smaller), and the few SLSN observed in
larger star-forming galaxies always occur close to the nuclei of their
hosts. Another type of peculiar and highly energetic supernovae are
the broad-line type Ic SNe (SN Ic-BL) that are associated with
long-duration gamma-ray bursts (LGRBs).  Also these have a strong
preference for occurring in small (SMC/LMC-like or smaller)
star-forming galaxies, and in these galaxies LGRBs always occur in the
brightest spots. Studies of nearby star-forming galaxies that are
similar to the hosts of LGRBs show that these brightest spots are
giant HII regions produced by massive dense young star clusters with
many hundreds of O- and Wolf-Rayet-type stars. Such dense young
clusters are also found in abundance within a few hundred parsecs from
the nucleus of larger galaxies like our own. We argue that the SLSN
and the SN Ic-BL/LGRBs are exclusive products of two types of
dynamical interactions in dense young star clusters. In our model the
high angular momentum of the collapsing stellar cores required for the
“engines” of a SN Ic-BL results from the post-main sequence mergers of
dynamically produced cluster binaries with almost equal-mass
components. The merger produces a critically rotating single helium
star with sufficent angular momentum to produce a LGRB; the observed
``metal aversion'' of LGRBs is a natural consequence of the model.  We
argue that, on the other hand, SLSN could be the products of
runaway multiple collisions in dense clusters, and we present (and
quantize) plausible scenarios of how the different types of
SLSNs can be produced.
\end{abstract}



\section{Introduction}

In the past decades several new and rare types of extremely bright and
peculiar supernovae have been discovered: 
\begin{itemize}
\item[(i)] The broad-line type Ic supernovae (abbreviated as SN Ic-BL)
  that are associated with long-duration GRBs (abbreviated as
  LGRBs). This was the first extremely bright type of supernovae
  discovered
  \citep{1998Natur.395..670G,2007MNRAS.375.1049W,2012Sci...337..932G,2012grb..book.....K}.
  They have no H and often also no He in their spectra and are
  characterized by extremely large outflow velocities ($\sim 40,000$
  km/s), implying very large kinetic energies ($\sim
  10^{52}$\,ergs). When they are associated with a LGRB (also a number
  has been discovered that are not, see below) they are related to the
  explosions of rapidly rotating almost pure CO-stars with masses $>
  5$\,\MSun, almost bare cores of originally very massive stars
  \citep{1998Natur.395..672I}, and the prototype SN1998bw/GRB980425
  ejected of order half a solar mass of $^{56}$Ni
  \citep{2011ApJ...740...41C}. Since the discovery of the LGRB-related
  supernovae of type Ic-BL, also non-GRB supernovae of this type have
  been discovered, in general in low-redshift galaxies \cite[$\langle
    z \rangle \sim 0.04$;][]{2012arXiv1211.7068G}. They are thought to
  have the same “central engines” that in LGRBs produce relativistic
  jets \citep{2010Natur.463..513S}; the jets are thought to be unable
  to penetrate the outer layers of the star, and to deposit their
  energy mostly inside the star, producing a SN Ic-BL
  \citep{2010Natur.463..513S,2010ApJ...725.1337L}.
\item[(ii)] The so-called super-luminous supernovae (SLSN), a new
  class of supernovae discovered with the recent large-scale surveys
  for transients. The several tens of extremely energetic and bright
  SLSNe that are now known have bolometric luminosities up to some 50
  times those of type Ia supernovae \citep{2012Sci...337..927G}. There
  are at least three classes of SLSN: the SLSN-I which lack hydrogen
  in their spectra, the SLSN-II which do have H in their spectra and
  the SLSN-R which have a long lightcurve tail powered by the
  radioactive decay of a very large amount of $^{56}$Ni, typically of order
  5\MSun\, \cite[for a review see][]{2012Sci...337..927G}.
\end{itemize}
Both the SLSN and the LGRBs (as well as their SN~Ic-BL counterparts)
have in common that: (1) They are very rare: the rate of LGRBs
($10^{-7}$\,Mpc$^{-3}$yr$^{-1}$) is some $10^3$ times lower than the
core-collapse SN rate \cite[the non-GRB SNIc-BL may be one to two
  orders of magnitude more common, e.g.\, see][but still
  rare]{2012arXiv1211.7068G}. The combined rate of the SLSN is of
order $10^{-8}$\,Mpc$^{-3}$yr$^{-1}$, some 10$^4$ times lower than the
core-collapse SN rate. These rates imply that a rare type of stellar
evolution is required to produce these events.  (2) Both the LGRBs and
SLSN occur almost exclusively in small star forming galaxies
(SMC/LMC-like or smaller; the same appears to be true for the non-GRB
SNIc-BL \citep{2012ApJ...759..107K}. \cite{2006Natur.441..463F} found
that in 41 out of 42 studied LGRBs the host is a small star-forming
galaxy. Only one LGRB host is a grand-design spiral galaxy, and it was
found that in their small hosts the GRBs fall on optically bright
spots. Studies of nearby small starburst galaxies show that such
bright spots are clumps of massive O- and Wolf-Rayet (WR) stars. For
example NGC 3125 has a number of such clumps with spectra that are a
mixture of O- and WR spectra \citep{2006MNRAS.368.1822H}. Studies of
these clumps show that such small galaxies may harbour as many as
$10^4$ O- and WR stars, which are concentrated in a small number (3 to
6) massive young star clusters, with masses of order
$10^5$\,\MSun\,each, and each containing often $> 600$
O-stars. 

The SLSN share the property of the LGRBs to occur almost exclusively
in small starburst galaxies. The only two SLSN-II that reside in
larger Milky Way-type galaxies were found very close to the nucleus of
their hosts \citep{2011ApJ...735..106D,2012Sci...337..927G}.
\cite{2012Sci...337..927G} remarks that this ``{\em suggests that to
  produce SLSN perhaps special conditions are required that are unique
  to this environment (e.g. circum-nuclear star-forming rings),
  somehow mimicking the conditions in star-forming dwarf galaxies.}''
Indeed, in the inner few hundred parsecs of the bulges of many larger
galaxies, nuclear starbursts are going on. Seyfert galaxies as well as
our own Galaxy are prime examples \citep{2008flhs.book.....C}. Within
100 pc from the centre of our Galaxy many massive young star clusters
are present, of which the Arches and Quintuplet clusters are key
examples. These clusters in our Galaxy's central region are massive,
but even more important (as we will argue in \S\,\ref{Sect:SC}) is
their very high stellar density, which in the core of Arches exceeds
$10^6$\,\MSun/pc$^3$. Such a high star density is also characteristic
of the young clusters in small star-forming galaxies like the LMC and
in the central regions of starburst galaxies like M82
\citep{2013ApJ...766...20L}.

The LMC cluster R136, in the 30 Doradus region, has a high
density. The reason why the clusters in small star-forming galaxies
reach such high star densities may be related to the power in the
turbulent velocity spectrum which leads to a shorter free-fall
timescale of the gas than in the disks of large galaxies
\citep{2011ApJ...741...10K}.  In the high density star forming
  regions the trubulent energy spectrum is a power law $E(k) \propto
  k^{-\gamma}$ in which $\gamma = 1.85\pm0.04$
\citep{2009ApJ...707L.153P} rather than the usual $\gamma \apgt 3$
\citep{2007ARA&A..45..565M} (with more power in the large scale gas
motion).  This is consistent with the results of hydrodynamical
simulations of the formation of star clusters in which a turbulent
velocity field with more power at small structures stimulate the
formation of dense clusters
\citep{2009A&A...504..883B,2009MNRAS.397..232B,2010MNRAS.404..721M,2012ApJ...761..156F}.

\begin{figure*}
\begin{center}
\includegraphics[width=0.8\linewidth]{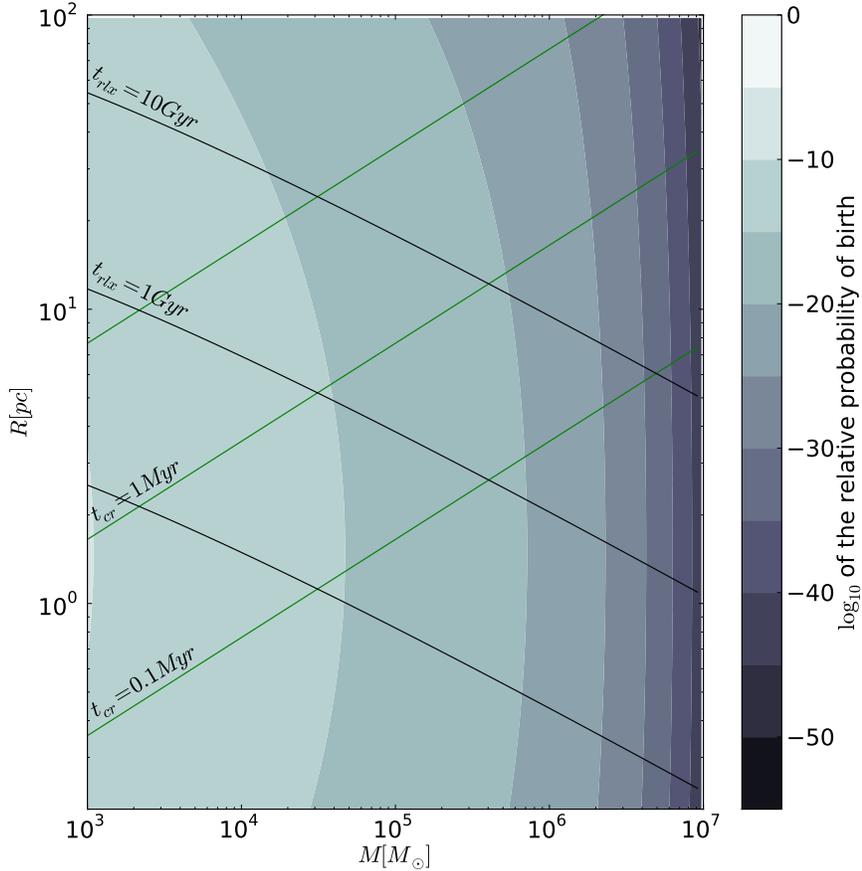}
\end{center}
\caption{The birth conditions (in mass and size) of Milky Way star
  clusters.  Gray shades (scale to the right) give the logarithm of
  the relative probability of birth. The birth probability density
  function is a convolution of the cluster initial mass function
  (Schechter function Eq.\,(\ref{Eq:Schechter}) with $M_* =2\cdot
  10^5$\,\MSun\, and $\beta=-3$.)  and the size distribution
  (log-normal with a mean of 5\,pc and a dispersion of 3\,pc).
The black curves (top left to bottom right) give the cluster 2-body
relaxation time, with $t_{\rm rlx} = 10$\,Gyr for the top curve down to
100\,Myr for the bottom curve.
The green curves (bottom left to top right) give the cluster crossing
time, with $t_{\rm cr} = 10$\,Myr for the top curve down to 0.1\,Myr
for the bottom curve.
\label{fig:MR_MWG_A}}
\end{figure*}

This effect can be observed in the population of star clusters in
nearby galaxies, by fitting their number $N$ to the Schechter 
function, which takes the form \citep{1976ApJ...203..297S}:
\begin{equation}
  NdM \propto M^\beta exp(M/M_*).
\label{Eq:Schechter}
\end{equation}
The distribution of the masses of young ($\aplt 10$\,Myr old) star
clusters in large quiescent galaxies, like M31, are best represented
by a Schechter function with a characteristic mass $M_* \simeq 2 \cdot
10^5$\,\MSun\, and with an exponential fall-off of $\beta \aplt -3$,
whereas for dwarf starburst galaxies and interacting galaxies like M51
$\beta \sim -2$ \citep{2010ARA&A..48..431P}.  In
Fig.\,\ref{fig:MR_MWG_A} (see also \S\,\ref{Sect:Origin}) we present
the probability density function of cluster birth mass and size. The
gray shades represent the convolution of the Schechter mass function
(Eq.\,\ref{Eq:Schechter}) with a log-normal distribution for the
cluster sizes.  For the former we adopted $\beta \aplt -3$ and $M_*
\simeq 2 \cdot 10^5$\,\MSun. For the log-normal size distbribution we
adopted a mean cluster radius of 5\,pc with a dispersion of 3\,pc,
which is consistent with the observed young ($\aplt 10$\,Myr) star
clusters in the local group \citep{2010ARA&A..48..431P}.  We speculate
that the dense torus of in-spiralling gas accumulating in the central
few hundred pc of the bulges of spiral galaxies may also have a
turbulent velocity structure, due to the high local star formation and
associated high supernova rate.

It therefore appears that both the LGRBs/SN Ic-BL and the SLSN solely
occur in regions of galaxies where very dense young star clusters are
present. This suggests that both types of objects could be the
products of evolutionary processes that are unique to dense massive
young star clusters, and do not occur anywhere else (this suggestion
for the LGRBs was casually made to one of us by S. Kulkarni in
2006). We present later in this paper (see \S\,\ref{Sect:Merger})
possible scenario of how this could come about and quantize the
effect in \S\,\ref{Sect:SC} and \S\,\ref{Sect:Discussion}. But before
that, in \S\,\ref{Sect:GRB}, we consider the boundary conditions set
by the observations, for models for producing LGRBs/SN Ic-BL.

\section{Conditions required for producing LGRBs/SN Ic-BL}\label{Sect:GRB}

The general concensus on the conditions required for producing a long
GRB is the collapse of a very rapidly rotating almost bare CO-core of
a massive star. There is strong observational evidence that the GRB is
produced by a narrowly collimated relativistic jet with a
Lorentz-factor of order $10^2$ to $10^3$
\citep{2006ARA&A..44..507W,2012Sci...337..932G}. The beam opening
half-angle of 5 to 10 degrees leads to a beaming fraction of order
0.003. There are two models for producing such jets: 
\begin{itemize}
\item[(i)] According to the ``{\em collapsar}'' model of
  \cite{1993ApJ...405..273W}, in which a very massive rapidly rotating
  core collapses to a black hole, the core has so much angular
  momentum that not all of the core matter can at once disappear into
  the black hole. Part of the core matter then temporarily forms a disk
  of nuclear matter around the black hole. Viscous and/or magnetic
  dissipation in this disk drive, in combination with frame dragging,
  a relativistic jet which results in the GRB
  \citep{1999ApJ...524..262M};
\item[(ii)] A model in which the very rapidly rotating core collapses
  to a strongly magnetized neutron star (“magnetar”) that is
  spinning with a period of order of a millisecond
  \citep{2011MNRAS.413.2031M,2011ApJ...726...90Z}. The spindown energy
  loss by magnetic dipole emission and the relativistic wind of such
  an extreme “pulsar” is so gigantic that it will spin down to a long
  period on a timescale of minutes to hours and produce energetic
  electromagnetically powered relativistic jets along the rotation
  axis, and also blows up the star in a SN Ic-BL \citep[][see also
    \cite{2010ApJ...717..245K,2010ApJ...719L.204W}]{2011MNRAS.413.2031M,2012Sci...337..932G}.
\end{itemize}

Both these models require that in order to produce a LGRB, the
collapsing massive CO core must have high angular momentum, in the
range $10^{16}$ to $10^{18}$ cm$^2$/sec
\citep{2006ARA&A..44..507W,2012grb..book.....K}. Since all SN Ic-BL
are thought to have the same central engine that produces highly
relativistic jets (see above), also the non-GRB supernovae of this
type must have collapsing CO-cores with the same high angular
momentum.  Two possible ways have been suggested for the core of a
massive star to obtain such high angular momentum, either: (a) very
low metallicity \cite[e.g.][]{2005A&A...443..643Y} or (b) evolution in
a close binary system
\cite[e.g.][]{2007PASP..119.1211F,2007Ap&SS.311..177V,2007ARep...51..308B,2008A&A...484..831D}. In
the first type of models it is argued that low metallicity gives weak
stellar winds – such that the winds do not carry off much angular
momentum and the star keeps high angular momentum throughout its
life. Its rapid rotation in these models, keeps the star completely
mixed, such that it evolves homogeneously and may in the end become a
rapidly rotating CO star that collapses
\citep{2005A&A...443..643Y,2006ApJ...638L..63L,2006A&A...460..199Y}.

The requirement of very low metallicity is fulfilled for many of the
host galaxies of LGRBs, but there are also several with almost solar
metallicity \citep{2007MNRAS.375.1049W,2012Sci...337..932G} and the
requirement of low metallicity must be reconsidered
\citep{2013arXiv1305.5165H}.  Simultaneously, the hosts of the non-GRB
supernovae of type Ic-BL tend to have low metallicity as well
\cite[e.g.][]{2012arXiv1211.7068G,2012ApJ...759..107K}; however, some
have metallicities as high as 1.7 to 3.5 times solar, while still
having the same central engine \citep{2010ApJ...725.1337L}. Therefore
it seems likely that low metallicity, although it appears to
facilitate the production of a LGRB and SN~Ic-BL, it is not the only
factor involved in the production of these phenomena, which require a
high angular momentum of the collapsing stellar core (see below).

\subsection{Gamma-ray bursts and SN~Ic-BL from regular v.s.\, dynamically formed binaries}

In close binary models -- involving late evolutionary phases of normal
massive binaries -- tidal forces keep the star in synchronous (rapid)
rotation \cite[e.g.\,][]{2007Ap&SS.311..177V}, or a rapidly rotating
merger product is produced \cite[``Helium merger
  GRB'',][]{2007PASP..119.1211F}.  The main argument against these
models, that involve the {\em regular} evolution of binaries that
started out as “normal” massive systems, is that such binaries are
found throughout the disks of all spiral galaxies. Therefore, if these
models would work, one would expect many LGRBs to be seen in disks of
spiral galaxies \cite[since a large part of the present-day star
  formation is thought to take place in these galaxies,
  e.g.\,see][]{2008flhs.book.....C}, contrary to what is
observed. Therefore, models based on {\em normal} massive binary
evolution cannot comply to the boundary conditions set by the
environments where LGRBs/SN Ic-BL are found.

The main open questions that then remain are: (1) why do LGRBs (and
other engine-driven SN Ic-BL) occur in small star-forming galaxies?,
and (2) why do LGRBs have a preference for low metallicities, while
not all SN~Ic-BL share this preference?

We propose that the answers to these questions are that the rapidly
rotating CO-cores required for the engines of SN~Ic-BL are solely
produced in mergers of a special type of binaries that result from
gravitational dynamical processes that occur only in dense young
massive star clusters. It turns out that dynamical processes tend to
produce close binaries with almost equal-mass components ($q\simeq1$).

In order for both stars to have a helium-burning helium core at the
time of the merger, the less massive star of the two should already
have left the main sequence (have exhausted hydrogen in its core)
while the more massive one should not yet have terminated core helium
burning. By performing a series of stellar evolution calculations
  using MESA \citep{2011ApJS..192....3P} from AMUSE framework
  \citep{2013CoPhC.183..456P} with solar ($\ZSun$) and sub-solar
  ($0.3\,\ZSun$) metallicity, we find that the two stars should not
  differ more than 20 per cent in mass, although the result is
  somewhat mass dependent allowing a larger mass difference for more
  massive star: the mean of the minimum mass ratio is $q\apgt
  0.87\pm0.07$ for solar metallicity and $q\apgt 0.89\pm0.06$ for
  subsolar. This poses a lower-limit for the required mass
ratio. The merger of such a binary produces a rapidly rotating massive
helium star, and we show that these stars at the time of core collapse
can still have the required high core angular momentum, and that their
event rates matches the observed rates of SN~Ic-BL. The winds of
helium stars (WR stars) carry off part of their original angular
momentum. We show that because of the metallicity dependence of the
wind mass-loss rates of Wolf-Rayet stars, at low metallicity high
final core angular momentum occurs over a much larger range of helium
star masses than at high metallicity. Therefore this model favors the
occurrence of LGRBs at low metallicity, but does not exclude high
metalicities. An additional factor favouring low metallicity is that
the helium merger product is more easily produced at low than at high
metallicity (see \S\,\ref{Sect:HeStar}).

\subsection{The evolution of the specific angular momentum of a 
            post-merger Helium star, resulting from an almost
            equal-mass binary}
\label{Sect:HeStar}

We used the models for the evolution of helium stars in the mass range
8 through 32\,\MSun\, of \cite{1978ApJ...219.1008A} and for larger
mass values the models of \cite[][models computed with more recent
  evolution codes produce very similal
  results]{1964ApJ...140..499D}. The total lifetimes, helium-burning
core masses, radii and luminosities of these models were adopted, and
for other helium star mass values in the range 8 to 100\,\MSun, these
quantities were calculated by logarithmic inter- and extra-polation
from these values as a function of the logarithm of mass. We
  assumed that at the time of the merger of the helium stars (cores of
  their progenitors) the merger product helium star is on the zero-age
  helium main sequence and is rotating with a break-up angular
  velocity $\Omega$. This is a natural consequence, because at the
  time of the merger the two helium cores are orbiting each other with
  keplerian velocities; the merger therefore results in a single
  helium star rotating with a keplerian equatorial angular velocity,
  which is the maximum possible ``break-up'' rate.  The assumption
  that they start on the zero-age helium main sequence implies that
  after their formation these stars have the longest possible
  lifetimes and therefore undergo the maximum possible amount of
  stellar-wind mass-loss (and thus maximum angular-momentum loss)
  that such a merger product can experience. (The real angular
  momentum loss of these merger products will therefore always be
  smaller than we calculate here, and their final core angular
  momentum will always be larger than we calculate here). For the
wind mass-loss rates we adopted the metal-dependent mass-loss rates of
WN-type Wolf-Rayet (WR) stars as given by \cite{2006A&A...460..199Y}:

\begin{equation}
  \log({\dot M_{\rm WR}/[\MSun/yr]}) = 
    -12.95 + 1.5 \log(L/\LSun)  + 0.86 \log(Z/\ZSun).
\label{Eq:WRwind}
\end{equation}
Here $Z$ and \ZSun\, indicate the star's metallicity and the Sun's
metallicity, respectively.

This mass-loss rate was adopted until the mass of the helium star had
been reduced to the mass of the convective burning core which it had
at the start of its evolution. Since in this core carbon is produced,
we assumed that from here on the WR star becomes a carbon type WC
star. For these stars we adopted the wind mass-loss rates for WC stars
given by \cite{2008flhs.book.....C}.

For $Z=\ZSun$, these rates are equal to the WN mass-loss rates given
by Eq.\,(\ref{Eq:WRwind}); for $Z=0.3\ZSun$ they are 3 times the rate
given by Eq.\,(\ref{Eq:WRwind}) for this metallicity, and for
$Z=0.1\ZSun$ they are 6 times the rate given by Eq.\,(\ref{Eq:WRwind})
for this metallicity.

We assumed the wind particles to carry off the angular momentum which
they had at the surface of the star, and since the bulk of the masses
of helium stars are convective, we assumed the stars to be rotating as
a solid body until the moment of helium exhaustion in the convective
core. After this, the contracting carbon-oxygen core will spin up, but
its rotation will be braked by coupling to the layers around it. As a
result it will lose part of the angular momentum it had at the time of
the helium exhaustion. After the end of carbon burning, at the time of
core collapse, it will still have a fraction $f$ of the specific
angular momentum which it had at helium exhaustion. We adopted the
values of $f$ as given by Yoon (2006, private communication), who
calculated the evolution of rotating helium stars with masses between
8 and 40\,\MSun\, using Spruit's (2002)\nocite{2002A&A...381..923S}
mechanism for core-envelope coupling. He found that the inner
3\,\MSun\, of the CO-cores of these stars at the moment of the core
collapse have retained a fraction $f$ of their initial specific
angular momentum which these had as a helium star in solid body
rotation.  These $f$-values are as follows: for $m_{\rm He} =
8$---16\,\MSun: $f= 0.20$; for $m_{\rm He} = 20\,\MSun$: $f= 0.40$;
for $m_{\rm He} = 25$\,\MSun\, $f = 0.65$ and for $M_{\rm He}=
40$\,\MSun: $f=0.75$.  For all masses $> 40$\,\MSun\, we adopted $f =
0.75$, and for other masses we estimated the $f$-values by logarithmic
interpolation as a function of logarithm of the mass.

The angular momentum of a star of mass $m$ and radius $r$, rotating at
angular velocity $\Omega$ is:
\begin{equation}
  J = mk^2r^2\Omega.
\end{equation}
Here $k$ is the radius of gyration of the star, which for helium stars
is given by \cite{1977ApJ...214L..19S}, as follows: for 8\,\MSun,
$k^2= 0.100$; for 16\,\MSun, $k^2= 0.115$ and for $M = 32$\,\MSun\,
and larger, $k^2 = 0.130$. For masses between 8 and 32\,\MSun\, the
$k^2$-values were obtained by logarithmic interpolation between the
above values.

The angular momentum loss rate is 
\begin{eqnarray}
  {dJ \over dt} &=& {d(mk^2r^2 \Omega) \over dt} \nonumber \\
                &=& mk^2r^2 {d\Omega \over dt}  +  k^2r^2\Omega {dm \over dt}.
\label{Eq:dJdt_A}
\end{eqnarray}
On the other hand,
\begin{eqnarray}
  {dJ \over dt} &=& r^2\Omega {dm \over dt}.
\label{Eq:dJdt}
\end{eqnarray}
In these equations, the radius $r$ of the star was assumed to be
constant during the WR phase, and the right-hand side of
Eq.\,(\ref{Eq:dJdt}) represents the angular momentum loss from the
stellar surface. For helium stars $\apgt 8$ solar masses the radii
indeed change little during the evolution; for the higher masses the
radii shrink somewhat in the course of helium burning, but as this
leads to a spin-up of the star, the angular momentum loss rate from
the surface will, in first approximation remain the same. We therefore
ignored the radius evolution of the helium stars. Combination of
Eqs.\,(\ref{Eq:dJdt_A}) and (\ref{Eq:dJdt}) then leads to:
\begin{equation}
  {d\ln \Omega \over dt} = \left( {1-k^2 \over k^2}\right) {d\ln m \over dt}.
\end{equation}
Integration yields:
\begin{equation}
  {\Omega_f \over \Omega_i} = \left( {m_f \over m_i}\right)^{1-k^2
    \over k^2},
\end{equation}
where the subscripts $i$ and $f$ indicate the initial and final
situations, respectively, and the exponent has a value between 7 and
9.  Although these exponents are large, our calculations below show
that for low matallicity the value of $m_i/m_f$ remains close to unity
over the range of masses.

\begin{figure*}
\begin{center}
\includegraphics[width=0.6\linewidth]{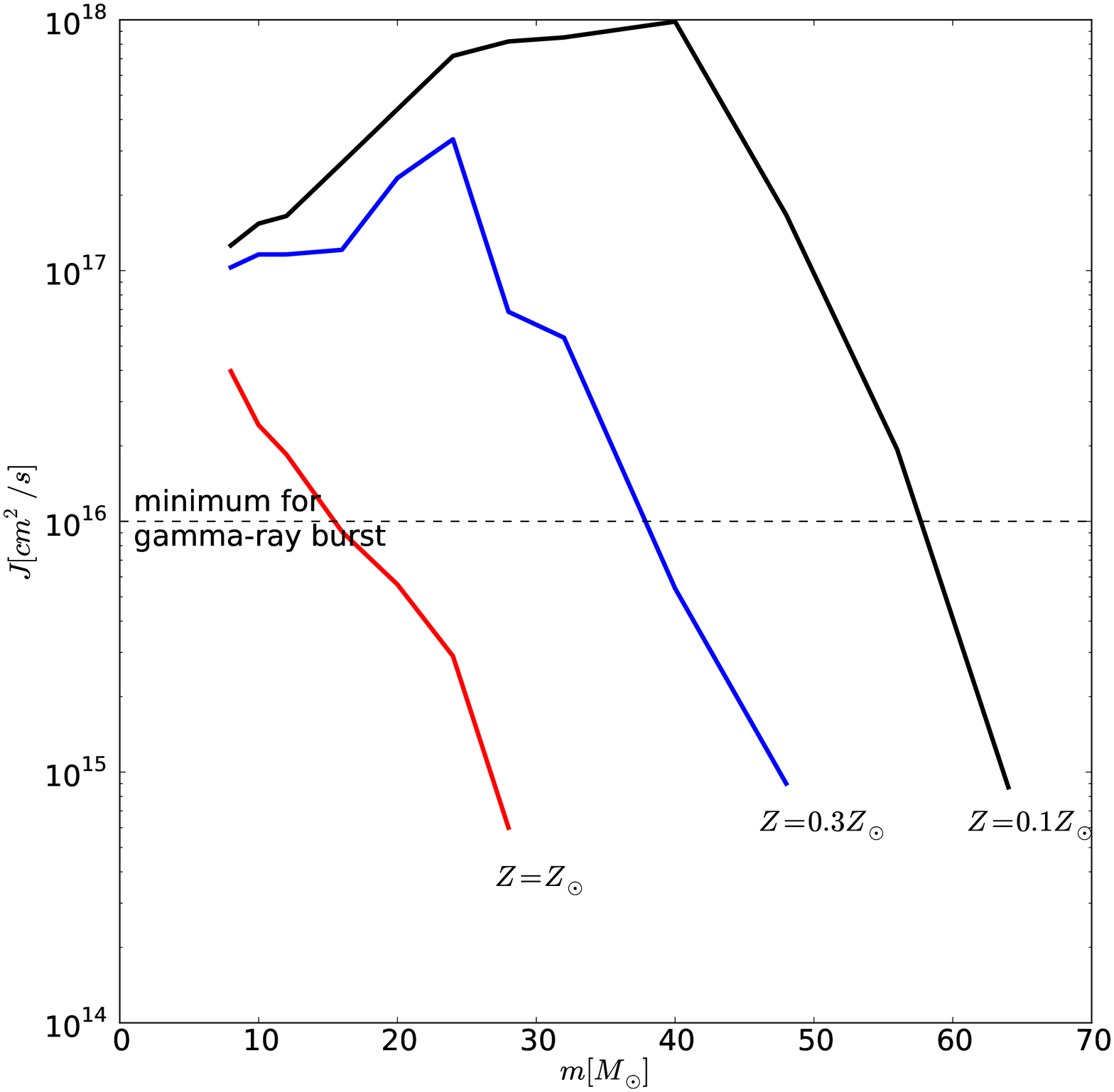}
\end{center}
\caption{Specific core angular momentum $J$ at the end of the
  evolution of a merger-produced helium star as a function of its
  zero-age helium-main sequence mass $m$, for solar metallicity,
  $Z=Z_\odot$ (red), $Z=0.3Z_\odot$ (blue) and for $Z=0.1Z_\odot$
  (black, as indicated).  The loss of angular momentum was calculated
  over the evolution of the Wolf-Rayet phase using the wind parameters
  by \cite{2006A&A...460..199Y} with an additional correction for the
  Carbon-WR phase. At birth, these merger-produced helium stars are
  assumed to spin at their break-up rotation rate. To produce a
  gamma-ray burst a minimum angular momentum of $J =
  10^{16}$\,cm$^2$/s is required
  \citep{2006ARA&A..44..507W,2012grb..book.....K,2010ApJ...716.1308L,2012ApJ...752...32W}.
\label{fig:AngularMomentum}}
\end{figure*}

We applied this equation, in combination with the above-given
$f$-values and the wind mass-loss rates and total lifetimes of the
helium stars calculated as defined above.  We started with helium
stars rotating at their break-up values after their formation by a
merger, and calculated the specific angular momentums $J_\star$ of
their collapsing cores at the end of the evolution, for three values
of the metallicity: $Z=\ZSun$, $Z = 0.3$\,\ZSun, and $Z =
0.1$\,\ZSun\, and for initial helium star masses in the range of
8---100\,\MSun. The resulting final specific angular momentum of the
collapsing CO cores for these three metallicities are depicted in
figure \,\ref{fig:AngularMomentum}. (The discontinuities in the slope
of the curves are in part caused by our interpolation methods combined
with the sudden jumps of the $f$-values, $k^2$-values and values of
some other quantities at certain helium star mass values as described
above.)

One observes that for Solar metallicity the range in zero-age masses
that have sufficient final core angular momentum for producing a
gamma-ray burst is considerably smaller than for lower metallicity,
but that high metallicities cannot be excluded for supernova Ic-BL
progenitors. For low metallicity the range in zero-age mass increases,
which is consistent with a higher proportion of gamma-ray bursts at
lower metallicity.

The curves show that for solar metallicity, the final specific angular
momentum of the cores are sufficient for producing LGRB only in the
helium-star mass range 8---16\,\MSun , for $Z=0.3$\,\ZSun the allowed
mass range is 8 to 38\,\MSun, and for $Z=0.1$\,\ZSun the range has
widened to 8 to 61\,\MSun.

We thus see that, in principle, Ic-BL and LGRBs can according to this
merger model, be produced at any metallicity, but that the mass range
in which such events can be produced is much larger at low metallicity
than at high metallicity. We speculate that this is the reason why
LGRBs have a preference for occurring at low metallicity, but are
still occasionally seen in a high-metallicity environment.

We now consider dynamical processes in dense young star clusters that
could produce these helium star merger and collision products that
could power the engine of a SN Ic-BL.

\section{Proposed merger scenarios for dynamically produced binaries, resulting in rapidly rotating collapsing CO cores.}\label{Sect:Merger}

Our scenario concerns the merger of a dynamically-produced binary,
consisting of two massive stars that at the time of the merger are in
or on their way to core helium burning.  To be simultaneously in this
phase, the two stars should at the outset not differ much in mass.  In
the case of low metallicity the stellar-wind mass loss during the
hydrogen-burning evolution of the stars will be small and the two
stars will still have hydrogen-rich envelopes when they merge. During
hydrogen shell burning these low-metalicity stars evolve to become red
supergiants with very large radii.  A binary contains insufficient
room for such stars and a common envelope will ensue in which the two
compact cores of the stars spiral towards each other and merge,
forming a helium core that rotates near break-up (i.e. with the
maximum possible angular momentum). During the common-envelope phase
the hydrogen-rich envelope is ejected, due to the release of a very
large amount of gravitational binding energy by the shrinking of the
binary orbit \cite[e.g.\,
  see][]{1984ApJ...277..355W,2013A&ARv..21...59I}.

In the high-metallicity case the two massive stars will at the time of
the merger already have lost their hydrogen-rich envelopes due to the
strong stellar winds, and have become Wolf-Rayet stars, but the
outcome of the merger will also be a critically rotating helium
star. Also such an object is expected to produce a SN Ic-BL. However,
to make the two stars merge at this phase requires an extra agent,
since such hydrogen-poor stars (Wolf-Rayet stars) have small radii and
are not expected to go into a common-envelope phase on their own
accord. To make them merge, a third companion to this binary is
needed, which through the \cite{1962AJ.....67..591K} effect, in which
the exchange of angular momentum between an inclined outer orbit and
the tight inner orbit drives the latter to extremely high ($\apgt
0.9$) eccentricity. This evolution is likely to result in an
off-center collision between the two WR stars, leading to a rapidly
rotating helium star.

Both such types of almost-equal-mass systems, without and with a third
companion are expected to be produced by dynamical interactions in
dense young star clusters, as numerical studies of star-cluster
evolution have shown
\citep{1996ApJ...467..359H,1999A&A...348..117P}. These studies show
that in dense young star clusters the most massive stars rapidly sink
to the cluster center, where they tend to form binary systems with
components that are very close in mass
\citep{2008MNRAS.384..376G}. Further dynamical interactions with
cluster stars and binaries may lead to the expulsion of such a massive
binary from the cluster, turning it into a runaway star
\citep{1988AJ.....96..222L,2011Sci...334.1380F}. The very massive
almost-equal-mass binary R145 near the LMC cluster R136 appears to be
precisely such a kicked-out runaway binary \citep{2013MNRAS.432L..26S}
(also the equal-mass close binary Y Cygni, which consists of two equal
mass B0 IV stars, in an eccentric 3-day orbit, is such a runaway star,
of lower mass, in our own galaxy; its cluster of origin is, however,
not known, \citep{1982ApJ...260..240G}). These ejected binaries tend
to be the ideal candidates for producing LGRB/SN Ic-BL, and our model
therefore predicts that LGRBs/SN Ic-BL can also be found
outside, though near (at a distance $300$---$10^3$\,pc) massive star
clusters. This appears indeed to be the case for some of the LGRBs/SN
Ic-BL, the prime example being GRB 980425/SN 1998bw
\citep{2006A&A...454..103H}.

\subsection{The link with dense young star clusters}\label{Sect:SC}

The above scenario only works for clusters for which a massive star
reaches the cluster core and pairs off in a binary before it leaves
the main sequence.  The most massive star $m_{\rm max}$ in a cluster
of mass $M$ is $m_{\rm max} \simeq 1.2 (M/\MSun)^{0.45}$\,\MSun\,
\citep{2004MNRAS.348..187W} with a maximum of about 150\,\MSun\,
\citep{1998ApJ...493..180M,2005Natur.434..192F}.  In our case,
however, the helium core of the most massive star, before merger,
should be $\aplt 30$\,\MSun\, (Fig.\,\ref{fig:AngularMomentum}), such
that the two cores together form a helium star $\aplt 60$\,\MSun. This
implies that the ZAMS hydrogen-rich progenitors should have been less
massive than about 61--68\,\MSun. (The range here reflects the
uncertainty in the moment when the common-envelope ensues which
translates in a range of core masses at the onset of Roche-lobe
overfow.) This corresponds to a cluster mass of 6200 to 7900 solar
masses.  In order to be able to form a binary the star has to sink
from the cluster virial radius $R$ to the cluster center within its
main-sequence life time.  This happens on a dynamical friction time
scale \citep[][here we adopted for the Coulomb logarithm $\log
  (\Lambda) = \log(0.1N)$]{1987gady.book.....B}:
\begin{equation}
    t_{\rm df} \simeq 2.2 {\rm Myr} 
                    \left({R \over {\rm pc}}\right)^{3/2} 
                    \left({M \over 10^4 \MSun} \right)^{1./2} 
                    \left({m_{\rm max} \over 150 \MSun} \right)^{-1}.
\label{Eq:tdf}
\end{equation}

By this time the most massive stars have sunken to the cluster center.
A single massive star, or one in a binary system, will upon arrival in
the central portion of the star cluster acquire a companion of similar
mass to form a binary or higher-order system
\citep{1996ApJ...467..359H,2008MNRAS.384..376G}.  A newly formed
binary will at first be rather wide, with a binding energy comparable
to the mean kinetic energy of the stars, or $\sim 1$kT.  Repeated
interactions with other cluster members drive the hardening of the
binary to $\apgt 100$\,kT.  In a number of cases, such a binary in the
cluster center may dynamically acquire a third companion, which later
in life, through the \cite{1962AJ.....67..591K} effect, may lead to a
collision of the inner pair.

In Fig.\,\ref{fig:MR_MWG} we present a number of observed star
clusters from the compilation of \cite{2010ARA&A..48..431P}.  The red
dashed curve indicates the cluster parameters, mass $M$ and
(virial/effective) radius $R$, for which the dynamical friction time
scale Eq.\,(\ref{Eq:tdf}) equals the main-sequence lifetime of the
most massive star. Clusters that are born with parameters below this
curve are prone to quick mass-segregation and form the relevant
population for supernovae producing type SN~Ic-BL progenitor binaries.

The young and dense galactic star cluster NGC3603 is in the regime for
this process, and may produce a supernova type Ic-BL. The cluster
contains a 3.77\,day double-lined eclipsing binary with a $116\pm
31$\,\MSun\, primary star NGC3603-A1 and a secondary star of
$89\pm16$\,\MSun\,\citep{2008MNRAS.389L..38S}, which could be a
prototypical example of a such a binary, although, just like in the
Quintuple cluster \citep{1999ApJ...514..202F}, its metallicity may be
too high to produce a LGRB (see Fig.\,\ref{fig:AngularMomentum}).
Several of the most massive stars in the Quintuplet cluster are known
to be binaries, but orbital parameters have not yet been determined
\citep{2012A&A...540A..14L}.  The central star cluster R136 in the 30
Doradus region of the large Magellanic cloud may be sufficiently dense
to producing a supernova type SN~Ic-BL, although the observed
parameters are controversial, in Fig\,\ref{fig:MR_MWG} we adopted
those reported by \cite{2013A&A...552A..94S}.  This cluster may have
ejected the object R144, which \cite{2011Sci...334.1380F} predicted to
be a massive binary. Recently \cite{2013MNRAS.432L..26S} identified
R144 as a $\aplt 370$\,day spectroscopic binary with a total mass of
$\sim 200$---300\,\MSun, confirming this earlier prediction.

The densest star clusters experience core collapse shortly after
birth. This happens in a small fraction of the two-body relaxation
time $t_{\rm cc} \sim 0.15t_{\rm rlx}$
\citep{2002ApJ...576..899P,2006ApJ...640L..39G}, as long as this time
scale does not exceed the main-sequence lifetime of the most massive
star, otherwise the cluster will expand due to copious stellar mass
loss.  Here the relaxation time of a cluster with effective (virial)
radius $R$ and mean stellar mass $\langle m \rangle \equiv M/N$ is:
\begin{equation}
    t_{\rm cc} \simeq 3.0{\rm Myr} \left( {M \over
      10^4\MSun} \right)^{1/2} \left( {R \over {\rm pc}} \right)^{3/2}
    \left( {\langle m \rangle \over \MSun} \right)^{-1}.
    \label{Eq:tcc}
\end{equation}
The red solid curve in Fig.\,\ref{fig:MR_MWG} indicates the cluster
parameters, mass $M$ and (virial) radius $R$, for which the
core-collapse time scale Eq.\,(\ref{Eq:tcc}) equals the main-sequence
lifetime of the most massive star.  

Clusters that experience core collapse before their most massive stars
have left the main-sequence are prone to many strong dynamical
interactions in the cluster center, and may experience a collision
runaway.  The collision rate during the time between core collapse and
the supernova explosion of the collision runaway product determines
the maximum mass of the latter.  \cite{2002ApJ...576..899P} estimated
the mass of a collision runaway as:
\begin{equation} 
    M_{\rm run} = 0.01 M \left(1 + {t_{\rm rlx} \over 100 {\rm Myr}} \right)^{-1},
\end{equation} 
with the additional requirement that $t_{\rm cc} < t_{\rm MS}(m_{\rm
  max})$.  The upper (left most) blue curve in Fig.\,\ref{fig:MR_MWG} 
indicates the cluster parameters for which a collision runaway leads to
a single object with a mass in excess of the most that of massive
main-sequence star ($M_{\rm run} \apgt 150$\,\MSun). Clusters to the
left of this curve but below the solid red curve will not grow a
massive runaway, but repeated interactions in the core may lead to the
ejection of the most massive binary, as a high-velocity
runaway. Although ejected, these binaries are still consistent with
the earlier discussed SN~Ic-BL/LGRB progenitors. We therefore
speculate that clusters born in this range of parameters are likely to
produce supernova type SN~Ic-BL that, by the time of the exploding
star is outside the cluster.  With a typical velocity of $\apgt
100$\,km/s and within $\sim 3$\,Myr to travel, the supernova type
SN~Ic-BL may occur $\apgt 300$\,pc from the cluster.

The young and dense Galactic star cluster Trumpler 14 is in the proper
regime of parameter space for producing a supernova type SN~Ic-BL by a
dynamically formed massive binary in its center.  Although the cluster
is still too young to have experienced core collapse a massive binary
is already present \citep{2009AJ....137.3358M}.  Based on our analysis
we expect that the binary in Trumpler 14 will eventually be ejected
from the cluster center and produce a supernova at some distance
away, but due to the high metallicity of Tr\,14, this explosion will
probably not resemble a LGRB/SN~Ic-BL.

According to our model each of these clusters are candidates for
producing a SN Ic-BL, each of which is expected to go off within the
next 3\,Myr totaling a rate of $\sim 1$/Myr, or $\sim 10^{-4}$ of the
type II supernova rate.  In Tab.\,\ref{Tab:rate} we calculate the rate
for supernova type SN~Ic-BL from the galactic star cluster population
and arrive at a theoretical upper limit of $1.5 \cdot 10^{-3}$ per
year.

\section{The origin of SLSNe}\label{Sect:Origin}

Dense clusters that are more massive than indicated by the left most
blue curve in Fig.\,\ref{fig:MR_MWG} but below the solid red curve are
prone to producing an unusually massive star via a collision runaway.
\citep{2002ApJ...576..899P}.  In relatively compact $R \aplt 0.4$\,pc
and low-mass star clusters $M \aplt 20,000$\,\MSun--30,000\,\MSun\,
the collision runaway product can reach a mass of 150\,\MSun\, to
260\,\MSun\, \citep{2007Natur.450..388P}.  \cite[According to][these
  limits are somewhat higher and occur between 250 and
  800\,\MSun.]{2008A&A...477..223Y} These stars collapse in a luminous
pair-instability supernova
\citep{1967ApJ...148..803R,2007A&A...475L..19L,2009msfp.book..209S,2012Natur.491..228C},
giving rise to a SLSN-R, because these supernovae produce large
amounts of $^{56}$Ni, as was proposed for SN2007bi by
\cite{2012MNRAS.423.2203P}.  The Arches star cluster is located in the
regime of forming a $\sim 170$\,\MSun\, collision runaway star, which
is in the range for leading to a pair instability supernova.

Over-plotted in Fig.\,\ref{fig:MR_MWG} (gray shades) is the
probability density function at which star clusters are born in the
Galaxy.  The gray shading is identical to that in
Fig.\,\ref{fig:MR_MWG_A}, but reproduced here to complement the
impression of cluster birth parameters (in gray) with the observed
population of star clusters.  For the size distribution of the
clusters we fitted the observed distribution of cluster sizes (taken
from Tabs 2, 3 and 4 of \cite{2010ARA&A..48..431P}) to a log-normal
distribution, which gave a satisfactory fit for a mean radius of 5\,pc
and a dispersion of 3\,pc.  For the initial mass function of young
clusters we adopted a Schechter function \citep{1976ApJ...203..297S}
with a minimum mass of $M = 500$\,\MSun\, and a characteristic mass of
$2 \cdot 10^5\,\MSun$ \citep{2009A&A...503..467L}.  The exponential
fall-off in the Schechter mass function for spiral galaxies is $\beta
\aplt -3$ (see Eq.\,\ref{Eq:Schechter}), whereas for dwarf
starburst galaxies it is $\beta \apgt -2$. This difference in shape of
the mass function, together with the adopted variation is the size
distribution gives rise to a dramatic difference in the densities for
these clusters (see Tab.\,\ref{Tab:rate}).

In sufficiently dense star clusters of $>30,000$\,\MSun\, the collision
runaway can grow to a mass $>260$\,\MSun\, (In Fig.\,\ref{fig:MR_MWG}
the area to the right of the right-most solid blue curve and below the
sold red curve). We speculate that these extremely massive stars
produce SLSN-I/II by collapsing to a black hole of intermediate mass
\citep{2009msfp.book..209S}.  The mass of the collision runaway can
reach values of up to a few
$10^3$\,\MSun\,\citep{2002ApJ...576..899P}.  By the time the star
experiences a supernova it has shed most of its mass again in a dense
stellar wind
\citep{2007ApJ...659.1576B,2008A&A...477..223Y,2009A&A...497..255G}
and it is uncertain how much mass eventually collapses to the black
hole \citep{2007ApJ...659.1576B}.  Integrating over the mass and size
distributions for star clusters we derive a rate of ${\cal R}_{\rm
  SLSN-I/II} \simeq 2.3 \cdot 10^{-7}$ for the Milky-Way population.
By adopting the same size distribution and mass distribution of star
clusters as we did before for the population of clusters in
blue-compact dwarf galaxies we arrive at a rate of ${\cal R}_{\rm
  SLSN-I/II} \simeq 3.4 \cdot 10^{-5}$, which is somewhat smaller than
the observed rate for combined types SLSN-I and SLSN-II.

\section{Discussion}\label{Sect:Discussion}

\begin{figure*}
\begin{center}
\includegraphics[width=0.8\linewidth]{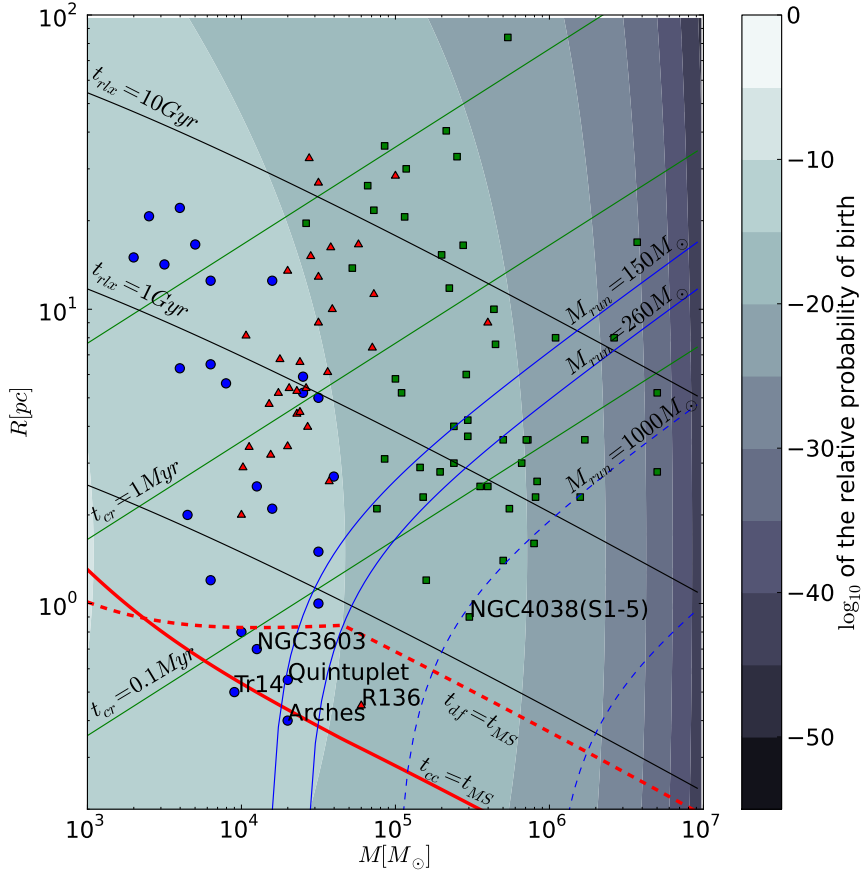}
\end{center}
\caption{Mass and effective radius for young clusters.  The
  gray-shades black and green curves are reproduced from
  Fig.\,\ref{fig:MR_MWG_A}.  The blue bullets, red triangles and green
  squares indicate the population of observed helium star clusters in
  the Milky Way, the local group and from nearby galaxies,
  respectively. The data was taken from table\,1 (blue), 2 (red) and 3
  (green) of \cite{2010ARA&A..48..431P}, except for the parameters for
  R136 (which are from \cite{2013A&A...552A..94S}), for Tr14 (from
  \cite{2012A&A...540A..57H}) and the Antennae cluster S1\_5
  \citep{2008A&A...489.1091M}.  Several of the densest clusters are
  identified by their common name.
The red dashed curve gives the radius of clusters for which the
dynamical friction time scale of the most massive star is the same as
its main-sequence lifetime.  Supernovae type Ic-BL are expected to
occur in clusters in the region below the dashed red curve, as long as
they have a relatively low mass of $\aplt 7900$\,\MSun.
The solid red curve indicates the radius of clusters for which the
core-collapse time scale is the same as the main-sequence lifetime of
the most massive star.
The blue curves (from bottom to the right side) indicate the mass of a
collision runaway that can form in these clusters.  Two of these
curves are identified as $M_{\rm run} = 150\,\MSun$, $M_{\rm run} =
260\,\MSun$.  (The two dashed blue curves indicate $M_{\rm run} =
1000\,\MSun$ and $M_{\rm run} = 10^4\,\MSun$.)  To the right of the
260\,\MSun\, massive runaway curve and below the solid red curve, we
expect star clusters to produce massive runaways that explode as
supernova type SLSN-I or SLSN-II.  Clusters in the region between the
two blue curves and below the solid red curve are expected to produce
a supernova type SLSN-R.
\label{fig:MR_MWG}}
\end{figure*}

We can calculate event rates for supernovae type Ic-BL, type SLSN-R
and SLSN-I/II by integrating the probability density function of star
cluster birth parameters and over galaxy types.  The integrated rates
for a large spiral galaxy and dwarf starburst galaxies are presented
in Tab.\,\ref{Tab:rate}. The supernova type Ic-BL are calculated by
integrating the area below the dashed red curve in
Fig.\,\ref{fig:MR_MWG}.  Because it is in our model the most massive
star in a cluster that pairs off and produces a supernova type Ic-BL,
we adopt an upper limit for the most massive star in the cluster, and
integrate up to that cluster mass.

According to our analysis presented in
Fig.\,\ref{fig:AngularMomentum}, the appropriate helium core mass for
each of the merging stars should be at most $\sim 8$\,\MSun, 16 and
30\,\MSun\, for $Z=\ZSun$, $Z=0.3\ZSun$ and $Z=0.1\ZSun$,
respectively.  Such core masses are reached in zero-age main-sequence
stars of at least 23--26\,\MSun, 42--48\,\MSun\, and 61--68\,\MSun.
This mass relates to the most massive star born in clusters, which
then should not exceed of 700--900\,\MSun\, 2700--3600\,\MSun\, and
6200--7900\,\MSun, for $Z=\ZSun$, $Z=0.3\ZSun$ and $Z=0.1\ZSun$,
respectively. The lower metalicities correspond to the higher mass
limits for the zero-age main-sequence stars and consequently also for
the upper limit in the cluster mass range; supernova type Ic-BL are
expected to occur in relatively low-mass ($\aplt 7900$\,\MSun) star
clusters.

In Tab.\,\ref{Tab:rate} we compare the relative rates for
LGRB/SN~Ic-BL as a function of metallicity with the metallicity
dependency in the observed rates, using statistics of LGRBs by
\cite[][see also
  \cite{2007MNRAS.375.1049W}]{2010AJ....140.1557L}. Although this
statistics contains only 14 LGRBs the number of low (with an oxygen
abundance of $12+log(O/H)<8.2$ counting 5 LGRBs), medium (7) and high
($12+log(O/H)>8.7$ with 2 LGRBs) metallicity pose an interesting
relation which can be compared with our model calculations. The total
relative rate for LGRB/SN~Ic~BL was fixed at $2\cdot 10^{-3}$
\citep{2006Natur.441..463F}.

The event rate for supernova type SLSN-I/II is calculated by
integrating the area below the solid red curve and to the right of the
right-most solid blue curve, and the type SLSN-R rate is obtained by
integrating between the two solid blue curves and below the solid red
curve.  We normalized to the supernova type~II rate by counting the
number of stars between 8 and 25\,\MSun, and we adopted a minimum
cluster mass of 150\,\MSun.  The relative rates for the various types
of supernovae are presented in Tab.\,\ref{Tab:rate}.

\begin{table*}[hb]
\caption{\label{Tab:rate} Event rates for families of supernovae.
  Observed rates (third column) for SLSNe are from
  \citep{2012Sci...337..927G}, and we determined the relative rate for
  LGRB/SN~Ic-BL from the statistics by \cite{2010AJ....140.1557L},
  from a sample of 14 LGRBs with a range of metallicities.  The
  subsequent three columns give the various rates from our model
  calculations.  All the rates are normalized to the core collapse
  supernova (type II) rate.  The best values are from our adopted
  Schechter mass function with exponential mass dependency of $\beta
  \apgt -2$ and with a log-normal size distribution with mean of
  $\langle r \rangle = 3$\,pc, which represents the star clusters in
  blue-compact dwarf galaxies. The Milky Way Galaxy fits best with
  $\beta \aplt -3$ and $\langle r \rangle = 5$\,pc.  The
  characteristic mass in the Schechter function in both cases is $M_*
  = 2\cdot10^5$\,\MSun.  The last column gives a combined rate
  assuming a relative ratio in starburst-to-quiescent galaxies of 1:10
  \citep{2013A&A...552A..44L}. }
\begin{tabular}{lllllll}
\hline
\hline
SN-type              & metallicity & observed         &model CBG         & model MWG       & combined \\
 &                   &&($\beta=-2$,      & ($\beta=-3$,    & 1:10 ratio \\
 &                   &&$\langle r \rangle =3$pc)      & $\langle r \rangle =5$pc)    &  \\
\hline
LGRB/SN Ic-BL  & $Z=0.1\ZSun$&$0.8 \cdot 10^{-3}$   &$(7.1\pm0.1) \cdot 10^{-3}$&$2.9 \cdot 10^{-3}$ &$3.3 \cdot10^{-3}$ \\
LGRB/SN Ic-BL  & $Z=0.3\ZSun$&$1.0 \cdot 10^{-3}$   &$(6.5\pm0.2) \cdot 10^{-3}$&$2.8 \cdot 10^{-3}$ &$3.1 \cdot10^{-3}$ \\
LGRB/SN Ic-BL  & $Z=\ZSun$   &$0.2 \cdot 10^{-3}$   &$(3.0\pm0.6) \cdot 10^{-3}$&$1.5 \cdot 10^{-3}$ &$1.6 \cdot10^{-3}$ \\
SLSN-I/II &$\forall Z$  &$1.7 \cdot 10^{-4}$ &$3.4 \cdot 10^{-5}$&$2.3 \cdot 10^{-7}$ &$3.4 \cdot10^{-6}$ \\
SLSN-R    &$\forall Z$  &  $2 \cdot 10^{-5}$ &$4.6 \cdot 10^{-5}$&$7.0 \cdot 10^{-7}$ &$4.6\cdot10^{-6}$ \\
\hline
\end{tabular}
\end{table*}


The rate of the various events for a galaxy similar to the Milky Way
are more than one order of magnitude lower than the observed rate or
the rate derived for compact dwarf galaxies.  The relative proportion
of star formation in these various types of galaxies may easily be an
order of magnitude, major spiral galaxies dominating this rate
\citep{2013A&A...552A..44L}. In that case, the rate for supernova type
Ic-BL may still be dominated by large spiral galaxies compared to compact
dwarf galaxies, but for the superluminal supernovae this does not pose
a discrepancy.

The difference in the observed rates of types SLSN-I/II compared to
SLSN-R is about an order of magnitude, whereas in our models they are
comparable. The relative ratio between SLSN-I/II and SLSN-R can easily
be tuned by moving the boundaries in runaway mass between producing a
SLSN-R and a SLSN-I/II.  Adopting a lower limit to the mass of the
collision runaway to produce a SLSN-I/II of $\sim 180$\,\MSun\,
(instead of 260\,\MSun) would solve this discrepancy.

We do not explicitly make the distinction between type SLSN~I and
type II, but derive the total rate. Upon each collision several
\MSun\, of hydrogen is injected into the collision runaway, but this
mass is blown away in the copious stellar wind in a few $10^4$\,
years.  A collision between the runaway and a hydrogen rich star
shortly before the supernova of the former was proposed by
\citep{2007Natur.450..388P} to explain the SLSN-II 2006gy which
occurred close to the nucleus of a large galaxy
\citep{2006CBET..644....1Q}.  The ratio between the timescale on which
fresh hydrogen is injected into the collision runaway and the time
required to deplete the newly acquired hydrogen envelope determines
the ratio of SLSN type II relative to SLSN type Is. The observed
comparable rates of type SLSN-II relative to SLSN-I is consistent with
this regime of collisional growth \citep{2002ApJ...576..899P}.  Both
the rates for SLSN-R and for SLSN-I/II increase if a higher proportion
of star clusters are born with high density, like is the case for
clusters in blue-compact dwarf galaxies compared to the Milky-Way.

The intermediate-mass black hole that forms through a SLSN-I/II
  is expected to be located in the dense core of a collapsed star
  cluster.  In the core of such a star cluster the intermediate mass
a black hole is likely to be accompanied by another star, or otherwise
it is likely to acquire one within a core relaxation time scale. The
orbital period of such a binary typically is in the range of 50\, to
500\,days \citep{2006MNRAS.370L...6P}.  the observational
  repercussions of a massive black hole that is orbited by another
  massive star are profound, and could be characterised by a peculiar
  x-ray emission. The companion eventually will leave the
  main-sequence upon which Roche-lobe overflow is likely to ensue.
Such a phase of mass transfers from the captured star to the
intermediate-mass black hole may lead to an ultra luminous x-ray
source, much like the observed systems M82 X-1
\citep{2001MNRAS.321L..29K}, NGC1313 X-2 \citep{2011AN....332..422Z}
and HLX-1 \citep{2012Sci...337..554W}, NGC5408 X-1
\citep{2009ApJ...706L.210S} and NGC7479
X-1\,\citep{2011MNRAS.418L.124V}.  The observed periodicity in M82 X-1
(62 days), NGC5408 X-1 (115days) and HLX-1 (388days) and their x-ray
fluxes are consistent with a cluster member being captured by an
intermediate mass black hole and feeding the latter via a dense
stellar wind or Roche-lobe overflow.

\acknowledgments

It is a pleasure to thank the anonymous referee for the constructive
remarks for improving the manuscript and Daniel Schearer and Nathan de
Vries for discussions. This work was supported by the Netherlands
Research Council NWO (grants \#643.200.503 [LGM], \#639.073.803 [VICI]
and \#614.061.608 [AMUSE]), by the National Science Foundation under
Grant No. NSF PHY11-25915 and by the Netherlands Research School for
Astronomy (NOVA).

\end{document}